\documentclass[reprint,superscriptaddress,aps,prl,longbibliography]{revtex4-2}

\usepackage[utf8]{inputenc}
\usepackage[T1]{fontenc}
\usepackage{amsmath}
\usepackage{amssymb}
\usepackage{graphicx}
\usepackage[capitalize, nameinlink]{cleveref}
\usepackage{subfigure}
\graphicspath{{figures/}}
\usepackage{adjustbox}
\usepackage{booktabs}
\usepackage{multirow}

\usepackage{color}
\usepackage{rotating}
\usepackage{ulem}

\newcommand*{\mo}{\mu_{0}}
\newcommand*{\lex}{\lambda_{ex}}

\newcommand*{\om}{\omega}

\begin{document}

\title{Generalized fractional approach to solving partial differential equations with arbitrary dispersion relations}

\author{Kyle Rockwell}
\author{Ezio Iacocca}
\affiliation{Center for Magnetism and Magnetic Nanostructures, University of Colorado Colorado Springs, Colorado Springs, CO 80918 USA}

\date{\today}

\begin{abstract}
Fractional calculus has been used to describe physical systems with complexity. Here, we show that a fractional calculus approach can restore or include complexity in any physical systems that can be described by partial differential equations. We argue that the dispersion relation contains the required information relating the energy and momentum space of the system and thus fully describes their dynamics. The approach is demonstrated by two examples: the Landau-Lifshitz equation in a 1D ferromagnetic chain, an example of a periodic crystal system with a bounded dispersion relation; and a modified KdV equation supporting surface gravity waves or Euler dispersion, an example of an unbounded system in momentum space. The presented approach is applicable to fluids, soft matter, and solid-state matter and can be readily generalized to higher dimensions and more complex systems. While numerical calculations are needed to determine the fractional operator, the approach is analytical and can be utilized to determine analytical solutions and investigate nonlinear problems.
\end{abstract}

\maketitle

The dynamics of physical systems are typically represented with partial differential equations (PDE). The notion of PDE models is traced to the rates of change involved in a system, so that time and space become coupled. Some PDEs are ``universal'' in the sense that they arise for a variety of physical systems under appropriate assumptions~\cite{AblowitzBook}, e.g., the Korteveg - de Vries (KdV) and nonlinear Schr\"{o}dinger (NLS) equations. Other PDEs are specific to physical systems such as Euler and Navier-Stokes equations for fluids and Landau-Lifshitz equation for magnetism. A third class of PDEs are fractional in the sense that the rates of change are not integer but fractional, i.e., are defined by fractional calculus~\cite{West2020}.

Fractional calculus formulations have been successful in describing the dynamics of several systems that are characterized by complexity. Physically, this implies non-local media and multiscale systems, including fractals~\cite{Failla2020} and resulting in anomalous diffusion~\cite{Chen2010}. Applications in image processing have been also identified~\cite{Yang2016} and realizations of fractional PDEs have been obtained experimentally~\cite{Liu2023}. Recently, it was shown that the fractional KdV and NLS are integrable, and support soliton solutions~\cite{Ablowitz2022}, opening opportunities for fractional nonlinear dynamics. However, such fractional PDEs are, to the best of our knowledge, constrained to an underlying fractal model. This means that media lacking complexity continues to be investigated with regular PDEs despite the fact that anomalous diffusion can nonetheless occur due to defects or impurities. Additionally, the PDE models used in physical systems are often the result of simplifications due to assumptions at different time and spatial scales that deliberately eliminate the system's complexity to make PDEs analytically and numerically tractable with known methods. This is a significant problem in material science since novel functional materials are increasingly dependent on surfaces and atomic structure~\cite{Liu2016,Manzeli2017}, making complexity a vital ingredient in their modeling and theoretical study.
Here, we demonstrate that fractional calculus can be invoked to restore the complexity of physical systems that exhibit non-trivial energy-momentum spaces. These arise in a variety of situations, e.g., approximations of atomic lattices in a continuum representation such as the Landau-Lifshitz equation~\cite{Rockwell2024}, high-order PDEs such as the Kawahara equation~\cite{Kawahara1971}, and model reductions of complex systems such as the Whitham equations~\cite{Whitham1974,Carter2018}. The fractional operator is conveniently solved in Fourier space by means of the Riesz definition~\cite{Riesz1948}. We provide two examples in which this method can be applied. First, we consider the quantum-mechanical dispersion of Landau-Lifshitz equation~\cite{Rockwell2024} which is bound in momentum space to the crystal's first Brillouin zone (FBZ). Second, we consider the KdV equation with an Euler dispersion which composes an unbounded momentum space.

\begin{figure}[t]
\centering
\includegraphics[width=3.3in]{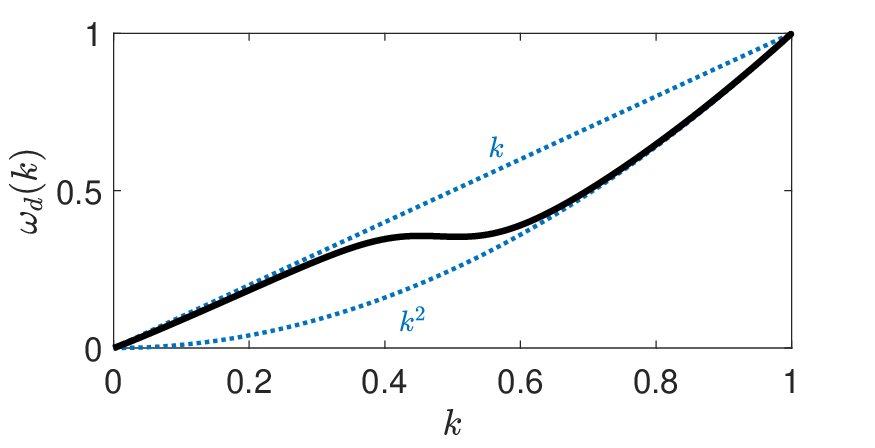}
\caption{Schematic illustration of a non-trivial dispersion relation $\omega_d(k)$ that can be defined between two models with integer differential operators. The transition can be described by a fractional differential operator.}
\label{fig1}
\end{figure}

To illustrate the rationale of our approach, consider a physical system whose dispersion relation is either experimentally determined or analytically found from a discrete system, $\omega_d(k)$. For example, we depict an arbitrary $\omega_d(k)$ in Fig.~\ref{fig1} with a solid black curve. This dispersion could be approximated by a linear dispersion when $k\rightarrow 0$ and a parabolic dispersion when $k\rightarrow 1$ with limits shown by dashed blue curves. The transition in between these limits would be difficult to tackle with a PDE and likely considered as different models. We propose here that $\omega_d(k)$ can be defined as a fractional, smooth transition from a linear to a parabolic dispersion. Physical examples where a similar transition can occur include the micromagnetic vs. the quantum-mechanical dispersion of magnons in ferromagnets~\cite{Rockwell2024}, the roton minimum in strongly interacting superfluids~\cite{Goldfrin2021}, and the dispersion relation from Euler equations compared to Serre and KdV equations~\cite{Carter2018}. Once the dispersion relation is described by a fractional operator, the resulting PDE could be solved either analytically or numerically with known methods, which is outside the scope of this letter.

To describe $\omega_d(k)$ fractionally, we invoke the Riesz definition~\cite{Riesz1948} of fractional operators. This is the most common approach for systems with bi-directional dispersion relations in Fourier space~\cite{Ablowitz2022,Laskin2002,Saichev1997}. The Riesz fractional derivative in the set of real numbers $\mathbb{R}$ of $\nu$-order is defined as,
\begin{equation}
    D_{x}^{\nu}\psi(x)= -\frac{D_{+}^{\nu}+D_{-}^{\nu}}{2\cos\left(\frac{\nu \pi}{2}\right)}\psi(x)\label{eq:ReRiesz}
\end{equation}
where the subscripts $+$ and $-$ correspond to the direction of approach for differentiating the function $\psi(x)$. The cosine in the denominator ensures even symmetry between forward and backward differentiation in the case where $\nu \ne 1$, and the minus sign guarantees that the integer second-order derivative is recovered as $\nu \rightarrow 2$. This produces a smooth and analytic connection from $d_{x}$ and $d^{2}_{x}$. The fractional order $\nu$ can also fall in the range $0<\nu<1$; this result is produced by integrating the operator. Although Eq.~\eqref{eq:ReRiesz} is defined in the range $0\le \nu < 1$, the Riesz fractional derivative can also reproduce the first order derivative by letting $\nu=1$ in Fourier space, whereas in $\mathbb{R}$, a Hilbert transform allows for $0\le \nu \le 2$~\cite{Riesz1948}.

The Fourier transform of Eq.~\eqref{eq:ReRiesz} is,
\begin{equation}
    \int_{-\infty}^{\infty} D^{\nu}_{x}\psi(x)e^{ikx}dx= -\frac{(ik)^{\nu}-(-ik)^{\nu}}{2\cos\left(\frac{\nu\pi}{2}\right)}\hat{\psi}(k)=-|k|^{\nu}\hat{\psi}(k)\label{eq:FTRiesz},
\end{equation}
where the subsequent modulus is a result of bidirectionally operating upon the Fourier kernel, allowing for the resolution of dispersion relations of even symmetry~\cite{Riesz1948}.  
It is clear from Eq.~\eqref{eq:FTRiesz} that the Fourier representation of the Riesz fractional differential operator is analogous to an integer differential operation in which $\nu$ generalizes the order of the differentiation. While the transform in Eq.~\eqref{eq:FTRiesz} is applied to the spatial components of $\psi(x,t)$, this approach could be also applied for the temporal case \cite{AlEssa2024}. Therefore, PDEs can be fully converted to a system-dependent fractional form.

Here, we constrain our analysis to PDEs of the form $\partial_\tau \psi(\tau,\xi)= D_{\xi}^{\nu}\psi(\tau,\xi)$, where $\tau$ and $\xi$ are dimensionless time and space, respectively. Therefore, the dispersion relation $\omega_{d}(k)$ can be written in the Riesz sense as
\begin{equation}
    |k|^{\nu(k)}=\om_{d}(k)\rightarrow \nu(k)= \frac{\ln{\om_{d}(k)}}{\ln{|k|}}.\label{eq:omRiesimple}
\end{equation}
The application of fractional calculus produces a unique outcome where the $\nu$ operator is directly proportional to the spatial and temporal elements of $\psi(x,t)$, thus producing an infinite set of case-dependent fractional solutions of various PDEs. We note that the approach can be immediately generalized to multiple spatial dimensions as well as higher-order time derivatives or fractional time operators.

An apparent problem is the singularity in Eq.~\eqref{eq:omRiesimple} as $k \rightarrow 1$ in dimensionless form. We provide two solutions to this issue. First, we employ a Taylor expansion of Eq.~\eqref{eq:omRiesimple} at $k+k_{0}$, with the expansion point depending on the system in question. {Because Eq.~\eqref{eq:omRiesimple} involves natural logarithms, $k_0$ must be chosen so that the singularity is present within the Taylor's expansion radius of convergence. This approach} bypasses the singularity and maintains the intuitive trend in $\nu(k)$, {e.g. between 1 and 2 in Fig.~\ref{fig1},} but at the cost of {numerical accuracy and high-order polynomial expansions}. For unbounded dispersion relations, the generated Taylor polynomial {can be} connected to the solutions from Eq.~\eqref{eq:nu_ck} by analytic continuation{, ensuring a smooth function in the full domain of $k$}.

A second approach is to introduce a function $\mathcal{C}(k)$, which produces a set of coefficients re-scaling the fractional operators to avoid the discontinuity. This approach has been employed for studying fractional diffusion~\cite{Leith2003}. {Therefore, Eq.~\eqref{eq:omRiesimple} is modified to}
\begin{equation}
  \nu(k)= \frac{\ln{\om_{d}(k)}}{\ln{|k|}+\mathcal{C}(k)}. \label{eq:nu_ck}
\end{equation}
{The limit of $\nu(k)$ at low wavevectors does not need to follow the integral differential model in this case and the function $\mathcal{C}(k)$ generally needs to be decaying so that the singularity effectively occurs at $k\rightarrow\infty$.}

To derive the set of possible functions $\mathcal{C}(k)$, {we interpret $\nu(k)$ as a classical path that can be generally parametrized by a Hamiltonian in polar coordinates}
\begin{equation}
   H = \frac{(d_kq)^2}{2}+r\cos({\theta}), \label{eq:Hamil}
\end{equation}
{where the first term is the conjugated kinetic energy and the second is the conjugated gravitational potential energy. In this representation, $q=r\theta$, $k=r\cos({\theta})$ and $\nu(k)=r\sin({\theta})$. In general, the path can be minimized by choosing $\mathcal{C}(k)$ by calculus of variations, resulting in an differential equation for $\mathcal{C}(k)$.} This approach is analytically exact, but at the expense of losing the physical interpretation of the fractional operator $\nu(k)$ and the complexity of solving the {differential equation for $\mathcal{C}(k)$. In fact, there are many, if not infinite, possible solutions for $\mathcal{C}(k)$ so that a guess must be chosen for the specific problem at hand. Based on the general requirements to avoid the discontinuity in Eq.~\eqref{eq:nu_ck}, i.e., a bounded and decaying $\mathcal{C}(k)$, we assume an exponential of the form}
\begin{equation}
   \mathcal{C}(k)=Ae^{-\beta k},\label{eq:cfun}
\end{equation}
{where $A,\beta>0$ are real numbers. Substituting Eq.~\eqref{eq:cfun} into Eq.~\eqref{eq:nu_ck} and imposing that the denominator does not cross zero, leads to $\ln A|k|e^{-\beta k}\leq0$. Solving for $\beta$ leads to the condition $\beta>Ae^{-1}$ while $A$ is solved from the limit $\lim_{k\rightarrow0}\nu(k)$.} This {approach} ensures that the denominator is never divergent. 

{For both methods, the} operators are constrained to reproduce the conserved $\om_{d}(k)$. Thus, the subsequent fractional group and phase velocities are also equivalent to the classical case; this difference consequentially {allows for the calculation of} the temporal components via {common integration methods} and the space-fractional components {by pseudospectral methods}. However, in systems of higher complexity, such as when non-conservative potentials are included, it is common to derive a modified dispersion relation and then relate it to the conservative $\om_{d}(k)$, leading to different group and phase velocities of the waves~\cite{Ablowitz2022}. 
\begin{figure}[t]
\centering
\includegraphics[width=3.3in]{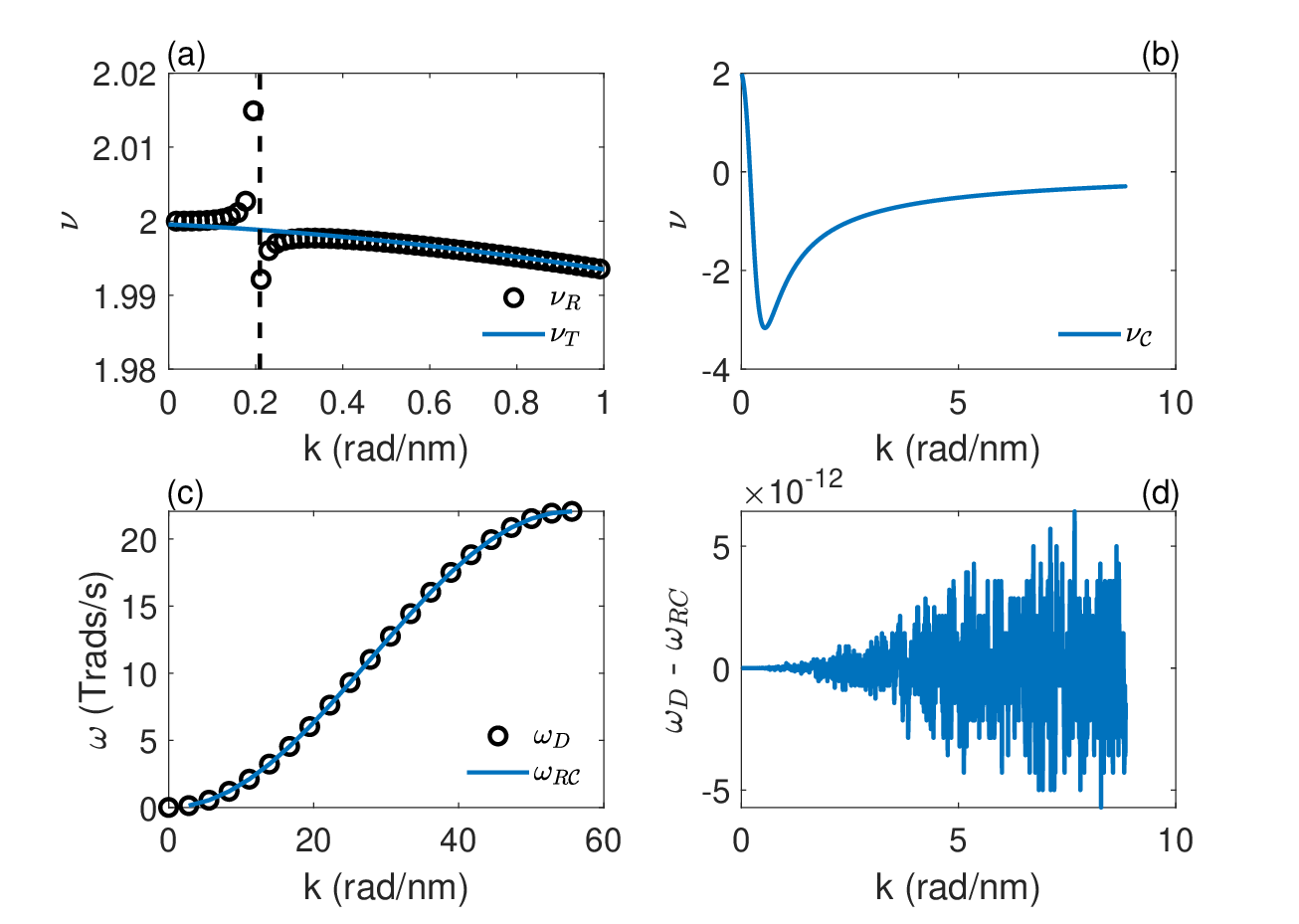}
\caption{(a) Calculation of the unmodified $\nu_{R}(k)$, shown in black circles, with the results from the Taylor expansion $\nu_{T}(k)$, shown as a blue curve, over a span of $1$~rad/nm; a black dashed line is used to mark the singularity at $0.2$~rad/nm. (b) Modified $\nu_{C}(k)$ with $\beta=3$. In this case, as $k\rightarrow 0$, the expected $\nu=2$ is restored. The nonlinear function $\nu_{C}(k)$ is not intuitive but describes the dispersion relation up to the FBZ, as shown in (c), where the black circles relate to the dispersion relation of magnons, and the blue line corresponds to the dispersion predicted by the modified $\om_{RC}(k)$. (d) Difference between $\omega_{D}(k)$ and the Riesz $\om_{RC}(k)$, where the error is in $\mathcal{O}(10^{-12})$.}
\label{fig2}
\end{figure}

{We now demonstrate how these methods recover the system's dispersion relation $\omega_d(k)$. We consider two cases: the bounded dispersion relation of the Landau-Lifshitz (LL) equations for ferromagnets~\cite{Landau1953} and the unbounded KdV equation modeled with an Euler-type dispersion relation~\cite{Carter2018}, a simplification of the Whitham equations.}

The LL equation describes the spatial and temporal evolution of the orientation of magnetic moments in a crystal structure, where $\mathbf{m}(x,t)$ is defined as {the magnetization vector, normalized to the saturation magnetization $M_s$ so that $|\mathbf{m}(x,t)|=1$.} The conservative portion of {the LL equation is simply the Larmor torque equation}
\begin{align}
    \partial_{t}\mathbf{m}(x,t)&= -\gamma\mo \mathbf{m}\times \mathbf{H}_\mathrm{eff},\label{eq:LL}
\end{align}
where $\gamma$ is the gyromagnetic ratio, $\mo$ the vacuum permeability, {and $\mathbf{H}_\mathrm{eff}$ is an effective field that contains the relevant physical interactions. Here, we are concerned exclusively with the exchange interaction. From the discrete Heisenberg Hamiltonian, the dispersion relation of the quanta of angular momentum, or magnons, is~\cite{White2007}}
\begin{equation}
    \om_{a}(k)= 2\gamma\mo M_s\left(\frac{\lex}{a}\right)^{2}\left[1-\cos(ka)\right],\label{eq:adisp}
\end{equation}
where $a$ is the crystal lattice constant {and $\lambda_\mathrm{ex}$ is the exchange length, a measure of the exchange stiffness in the system}. While Eq.~\eqref{eq:adisp} is accurate for {all magnons within the first Brillouin zone (FBZ), $0\leq k\leq\pi/a$, analytical and numerical work often use its long-wave approximation, reducing the system to the well known $k^2$ dispersion in ferromagnets. However, it has been recently shown that such approaches are inaccurate when describing far-from-equilibrium dynamics~\cite{Rockwell2024} and solitons with sharp profiles~\cite{Estiphanos2024} and the dispersion relation of Eq.~\eqref{eq:adisp} must be used instead.} Since $\om_{a}(k)$ is bidirectional, {the present} method guarantees that only magnons with the correct energy and momentum are capable of propagating in the system.

{Setting the exchange effective field to $\mathbf{H}_\mathrm{eff}=M_s D_{x}^\nu\mathbf{m}$ in Eq.~\eqref{eq:LL}, we apply Eq.~\eqref{eq:omRiesimple} to obtain}
\begin{align}
    \nu(k)&=\frac{2\ln\left[\left(\frac{\lex}{a}\right)^2(1-\cos(ka))\right]}{\ln[\lex k]}.\label{eq:nuLL_discont}
\end{align}

{In Eq.~\eqref{eq:nuLL_discont}, space is scaled by $\lambda_\mathrm{ex}$. Here, $\lim_{k\rightarrow0}\nu(k)=2$ as expected for the long-wave dispersion relation, i.e., expansion of Eq.~\eqref{eq:adisp}. The singularity occurs for the fractional operator $\nu(k)$} if $k=\lex^{-1}=0.2$~rad/m, {as shown by black circles in Fig.~\ref{fig2}(a) and the vertical dashed black line indicating the location of the singularity.}

{The first solution to this issue is using Taylor expansion. We chose $k=\pi/(2a)$, half the FBZ, to expand about. The solution in the vicinity of the singularity is shown in Fig.~\ref{fig2}(a) by a solid blue curve. The second solution is to implement the} decaying dimensionless function of coefficients, $\mathcal{C}(k)$. Thus, in this case, Eq.~\eqref{eq:nu_ck} becomes
\begin{equation}
    \nu(k)= \frac{2\ln\left[\left(\frac{\lex}{a}\right)^2(1-\cos(ka))\right]}{\ln[\lex k]+\beta k},\label{eq:LL_nu_ck}
\end{equation}
{where we set $A=1$. The singularity is removed if $\beta>\lex e^{-1}\approx 1.83$. Choosing $\beta=3$ we obtain the fractional operator $\nu(k)$ shown in Fig.~\ref{fig2}(b).} As previously stated, the resulting operators lose physical intuition, {as can be seen by $\nu(k)$ becoming negative, thus implying an integration. Nevertheless, we maintain the low-wavenumber limit $\nu(k)=2$ exactly. The dispersion relation is then recovered throughout the FBZ by computing $\omega_{RC}(k)=|\lambda_\mathrm{ex}e^{-3k}k|^{\nu(k)}$. This is shown in Fig.~\ref{fig2}(c) using parameters for permalloy, $M_s=790$~kA/m and $\lambda_\mathrm{ex}=5$~nm. The quantum-mechanical dispersion relation is shown by black circles while the fractional dispersion is shown by a solid blue curve. The computation is essentially exact, as seen by the error on the order of $10^{-12}$ shown in Fig.~\ref{fig2}(d).}

\begin{figure}[t]
\centering
\includegraphics[width=3.3in]{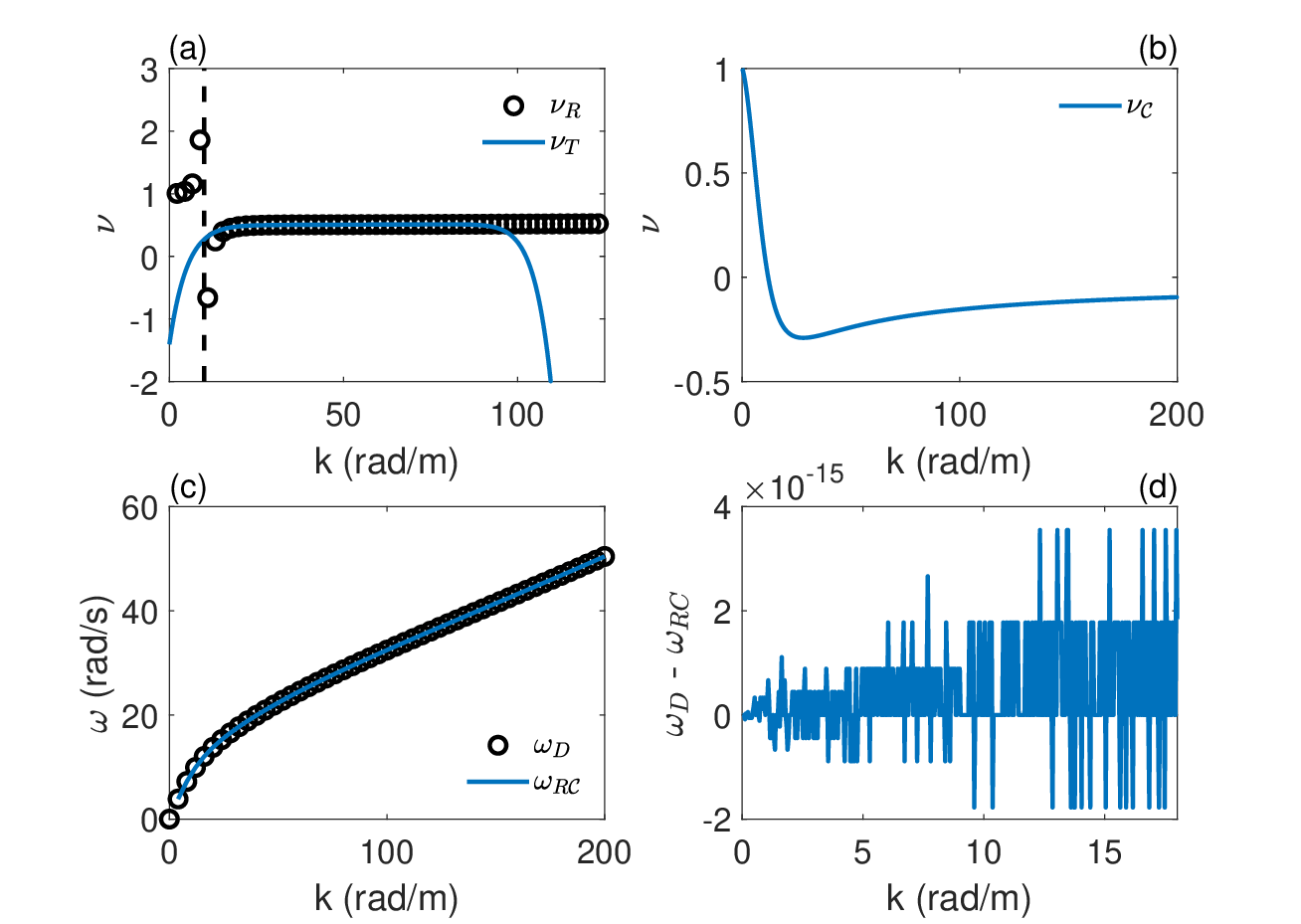}
\caption{(a) Calculation of the unmodified $\nu_{R}(k)$, shown in black circles, with the results from a Taylor expansion, at $k_{0}=5/h_{0}$~rad/m, labeled as $\nu_{T}(k)$, and shown as a blue curve, over a span of $125$~rad/m. (b) Modified $\nu_{C}(k)$ with $\beta=0.1$. In this case, as $k\rightarrow 0$, the expected $\nu=1$ is restored. $\nu_{C}(k)$ continues to produce the predicted dispersion relation as $k\rightarrow \infty$. (c) The range of the Euler dispersion relation up to $k=200$~rad/m $\om_{D}(k)$ is shown via black circles, where the blue curve corresponds to the angular frequency predicted by modified $\om_{RC}(k)$. (d) Difference between the Euler $\omega_{D}(k)$ and the Riesz $\om_{RC}(k)$, where the error is in $\mathcal{O}(10^{-15})$.}
\label{fig3}
\end{figure}
We now apply our approach to a modified form of the KdV equation~\cite{Carter2018}, where its dispersion relation has been replaced with that of the Euler equation,
\begin{equation}
\om_{E}(k)= \pm \sqrt{k\left(g+\tau k^{2}\right)\tanh{(kh_{0})}},\label{eq:omE}
\end{equation}
where, $g$ is gravitational acceleration, $\tau$ is the surface tension, and $h_{0}$ is the unperturbed depth of the water waves.
Again, $\om_{E}(k)$ is bidirectional, meeting the Riesz condition for $\nu(k)$. {This is an example of an unbounded dispersion relation insofar as the transition from fluid dynamics to molecular dynamics is assumed to occur at $k\rightarrow\infty$. From} Eqs.~\eqref{eq:omRiesimple}-\eqref{eq:nu_ck} and scaling time by $\sqrt{h_0/g}$ we obtain,
\begin{equation}
    \nu(k)=\frac{\ln\left[h_0k(1+\tau k^{2}/g)\tanh(kh_{0})\right]}{2\ln\left[h_0 k\right]},\label{eq:nu_Eu_discont}
\end{equation}
where the spatial scaling factor $h_0$ ensures that $\nu(k)$ is dimensionless and $\lim_{k\rightarrow0}\nu(k)=1$. {Following Ref.~\cite{Carter2018}, we set parameters $\tau=72.86\times10^{-6}$~m$^{3}$/s$^{2}$, $h_{0}=0.1$~m, and $g=9.81$~m/s$^{2}$. The discontinuity is located at $k=h_0^{-1}= 10$~rad/m. The scaled form of the fractional operator is,}
\begin{equation}
    \nu(k)=\frac{\ln\left[h_0k(1+\tau k^{2}/g)\tanh(kh_{0})\right]}{2\ln\left[h_0k\right] -\beta k},\label{eq:nu_Eu_ck}
\end{equation}
where, $\beta > h_0e^{-1}~\approx 0.037$~m. For our analysis, we set $\beta=0.1$.

Fig.~\ref{fig3}(a) shows the evolution of the fractional operator, $\nu_{R}(k)$, predicted by Eq.~\eqref{eq:nu_Eu_discont} in black circles, and the results from a Taylor expansion, $\nu_{T}(k)$, at $k_{0}=5/h_{0}$~rad/m, presented as a blue curve. Here, the Taylor expansion as chosen describes also the discontinuity thus failing to reproduce the physical limit when $k\rightarrow0$. In addition, as $k\rightarrow 100$~rad/m, $\nu_{T}(k)$ diverges from the expected result {as the radius of convergence is reached. In other words,} Due to the Euler dispersion relation being unbounded, it is impossible for the resulting Taylor polynomials to maintain the trend in $\nu_{R}(k)$ as $k\rightarrow\infty$; this issue can be solved by analytic continuation from $\nu_{T}(k)$ to $\nu_{R}(k)$.
 
{Use of} the scaled $\nu_{C}(k)$, shown in Fig.~\ref{fig3}(b) via a blue curve, accurately predicts $\nu=1$ as $k\rightarrow 0$. We constrained the $k$-axis to the same domain as in Fig.~\ref{fig3}(a) in order to exemplify how $\nu_{C}$ is continuous through the singularity. For Fig.~\ref{fig3}(c), we extended the $k$ domain to analyze the angular frequency, $\om_{RC}(k)$, shown via blue line and obtained from calculating Eq.~\eqref{eq:nu_Eu_ck} compared to Eq.~\eqref{eq:omE} shown via black circles. {Both dispersion relations are in excellent agreement,} with Fig.~\ref{fig3}(d) showing the difference in angular frequency on the order of $10^{-15}$. Thus, we consider the solution to be numerically exact.   

In summary, we have presented a fractional calculus approach to include or maintain complexity into PDEs. The approach is general in scope and it is shown to be applicable to both bounded and unbounded problems. This approach can be easily integrated into pseudospectral methods and take advantage of calculations in Fourier space. Analytical work can be performed with this formalism, insofar as the expression for the dispersion relation is tractable. It is expected that Fourier methods can be used to determine linear and nonlinear solutions to a variety of physical systems where periodic solutions can be enforced. Thus, the presented formulation can be easily extended to inertial systems, such as mechanics and antiferromagnets~\cite{Baltz2018}. In addition, it can be argued that experimental determination of dispersion relations can be used to inform the fractional PDE, thus completely describing the system in the presence of irregular structures.

\section*{Acknowledgments}

This work was supported by the U.S. Department of Energy, Office of Basic Energy Sciences under Award Number DE-SC0024339. The authors are thankful to Prof. Lincoln Carr for fruitful discussions.

\end{document}